\documentclass[10pt,conference]{IEEEtran}
\IEEEoverridecommandlockouts

\usepackage{cite}
\usepackage{amsmath,amssymb,amsfonts}
\usepackage{algorithm}
\usepackage{algpseudocode}
\usepackage{textcomp}
\usepackage{xcolor}
\usepackage{enumitem}
\usepackage{listings}
\usepackage{float}
\usepackage{graphicx}
\usepackage{enumitem}
\usepackage{adjustbox}
\usepackage{tcolorbox}
\usepackage{svg}
\usepackage{balance}
\usepackage{booktabs}
\usepackage{url}
\usepackage{algorithm}
\usepackage{algpseudocode}
\usepackage{amsmath,amsfonts}
\usepackage{graphicx,calc}
\usepackage{textcomp}
\usepackage{xcolor}
\usepackage{listings}
\usepackage[utf8]{inputenc}
\usepackage[flushleft]{threeparttable}
\usepackage{balance}
\usepackage{wrapfig}
\usepackage{textcomp}
\usepackage{mathtools,amsthm}  
\usepackage{tcolorbox}
\usepackage{xstring}
\usepackage{comment}
\usepackage{multicol}
\usepackage{tikz}
\usetikzlibrary{calc,arrows.meta}
\usepackage{xspace}
\usepackage{tcolorbox}
\usepackage{booktabs}
\usepackage{hyperref}
\usepackage{soul}

\newlength\myheight
\newlength\mydepth
\settototalheight\myheight{Xygp}
\newcommand{\bettersim}{{\raise.17ex\hbox{$\scriptstyle\sim$}}}
\lstdefinestyle{codingstyle}{
  basicstyle=\ttfamily,
  numbers=left,
  numberstyle=\tiny,
  backgroundcolor=\color{lightgray!20}, 
  frame=single,
  rulecolor=\color{gray!50}, 
  breaklines=true,
  captionpos=b,
  xleftmargin=1.5em,
  xrightmargin=1.5em,
  showstringspaces=false,
}
\usepackage{enumitem}
\newcommand{\squishlist}{
	\begin{itemize}[noitemsep, nolistsep, leftmargin=*]
		\setlength{\itemsep}{-0pt}
	}
	\newcommand{\squishstart}{
		\begin{itemize}
		}
		\newcommand{\squishend}{
		\end{itemize}
	}
	\newcommand{\PP}[1]{
		\vspace{3px}
		\noindent{\bf \textsc{\IfEndWith{#1}{.}{#1}{#1.}}}
	}
	\newcommand{\PPP}[1]{
		\vspace{2px}
		\indent{\it \IfEndWith{#1}{.}{#1}{#1.}}
	}
	
	\newcommand{\commentt}[1]{}

	\graphicspath{{images/}}

\newlength{\textfloatsepsave}
\usepackage{listings}
\usepackage{xcolor}

\colorlet{punctcolor}{red!60!black}
\colorlet{desccolor}{green!60!black}
\colorlet{altcolor}{blue!60!black}
\colorlet{kwcolor}{teal!60!black}
\definecolor{keywordcolor}{rgb}{0.13, 0.29, 0.53}

\begin{document}

\title{AutoRestTest: A Tool for Automated REST API Testing Using LLMs and MARL}
\author{
    \IEEEauthorblockN{Tyler Stennett\IEEEauthorrefmark{1}\textsuperscript{$1$}, Myeongsoo Kim\IEEEauthorrefmark{1}\textsuperscript{$2$}, Saurabh Sinha\IEEEauthorrefmark{2}\textsuperscript{$3$}, Alessandro Orso\IEEEauthorrefmark{1}\textsuperscript{$4$}}
    \IEEEauthorblockA{\IEEEauthorrefmark{1}Georgia Institute of Technology, Atlanta, Georgia, USA\\
    Email: \{tyler.stennett\textsuperscript{$1$}, mkim754\textsuperscript{$2$}\}@gatech.edu, orso@cc.gatech.edu\textsuperscript{$4$}}
    \IEEEauthorblockA{\IEEEauthorrefmark{2}IBM Research, Yorktown Heights, New York, USA\\
    Email: sinhas@us.ibm.com\textsuperscript{$3$}}
}

\maketitle

\begin{abstract}
As REST APIs have become widespread in modern web services, comprehensive testing of these APIs is increasingly crucial. Because of the vast search space of operations, parameters, and parameter values, along with their dependencies and constraints, current testing tools often achieve low code coverage, resulting in suboptimal fault detection. To address this limitation, we present AutoRestTest, a novel tool that integrates the Semantic Property Dependency Graph (SPDG) with Multi-Agent Reinforcement Learning (MARL) and large language models (LLMs) for effective REST API testing. AutoRestTest determines operation-dependent parameters using the SPDG and employs five specialized agents (operation, parameter, value, dependency, and header) to identify dependencies of operations and generate operation sequences, parameter combinations, and values. Through an intuitive command-line interface, users can easily configure and monitor tests with successful operation count, unique server errors detected, and time elapsed. Upon completion, AutoRestTest generates a detailed report highlighting errors detected and operations exercised. In this paper, we introduce our tool and present preliminary findings, with a demonstration video available at \href{https://www.youtube.com/watch?v=VVus2W8rap8}{https://www.youtube.com/watch?v=VVus2W8rap8}.
\end{abstract}


\section{Introduction}
REpresentational State Transfer (REST) APIs serve as the backbone of modern web services, with nearly 90\% of developers engaging with APIs and approximately 86\% of these APIs employing the REST architecture. By emphasizing a lightweight design and scalability, REST APIs seamlessly connect software systems through standard web protocols, such as the HyperText Transfer Protocol (HTTP). This approach supports a client-server architecture that effectively separates responsibilities while enhancing the efficiency and maintainability of web services~\cite{richardson2013restful}.

The critical role of REST APIs in web service interactions has sparked significant interest in automated techniques for testing these APIs, particularly following the introduction of the Open API Specification (OAS)~\cite{kim2022automated}, and many researchers and practitioners have proposed a variety of techniques in this space. 

For instance, RESTler employs search algorithms (BFS, DFS, and Random Walk), while EvoMaster~\cite{arcuri2019restful} utilizes evolutionary algorithms. For another example, MoRest~\cite{morest} manages operation dependencies through a dynamically updated RESTful Property Graph (RPG) derived from the OAS. Reinforcement learning has also been applied in this domain, with ARAT-RL~\cite{kim2023reinforcement} and DeepREST~\cite{corradini2024deeprest} optimizing parameter selection and search space exploration. More recently, large language models have been leveraged to further enhance testing: LlamaRestTest~\cite{kim2025llamaresttesteffectiverestapi} focuses on value generation and inter-parameter dependencies using language models, while NLP2REST~\cite{kim2023enhancing} and RESTGPT~\cite{kim2023leveraging} enhance the OAS by extracting actionable rules from natural language descriptions.

While existing research has addressed individual challenges of REST API testing, such as parameter generation, inter-parameter dependencies, and operation dependencies, no comprehensive technique has yet tackled these challenges together effectively. To address this limitation, in this paper we introduce AutoRestTest, a tool that combines graph-based dependency modeling, Large Language Models (LLMs), and Multi-Agent Reinforcement Learning (MARL) to perform REST API testing holistically. Key features of AutoRestTest include:

\begin{itemize}
\item The \textbf{Semantic Property Dependency Graph (SPDG)} enables the use of a dependency agent to reduce the search space and explore API dependencies efficiently.
\item \textbf{REST Agents} effectively identify headers, operations, parameter combinations, and their corresponding values.
\item An \textbf{LLM} model creates realistic inputs for both value and header generation.
\item A friendly \textbf{User Interface} provides a command line interface (CLI) alongside detailed reports that highlight the results of the tool, including detected internal server errors.
\end{itemize}

\section{AutoRestTest}

\begin{figure*}[t]
\centering
\includegraphics[width=.9\textwidth]{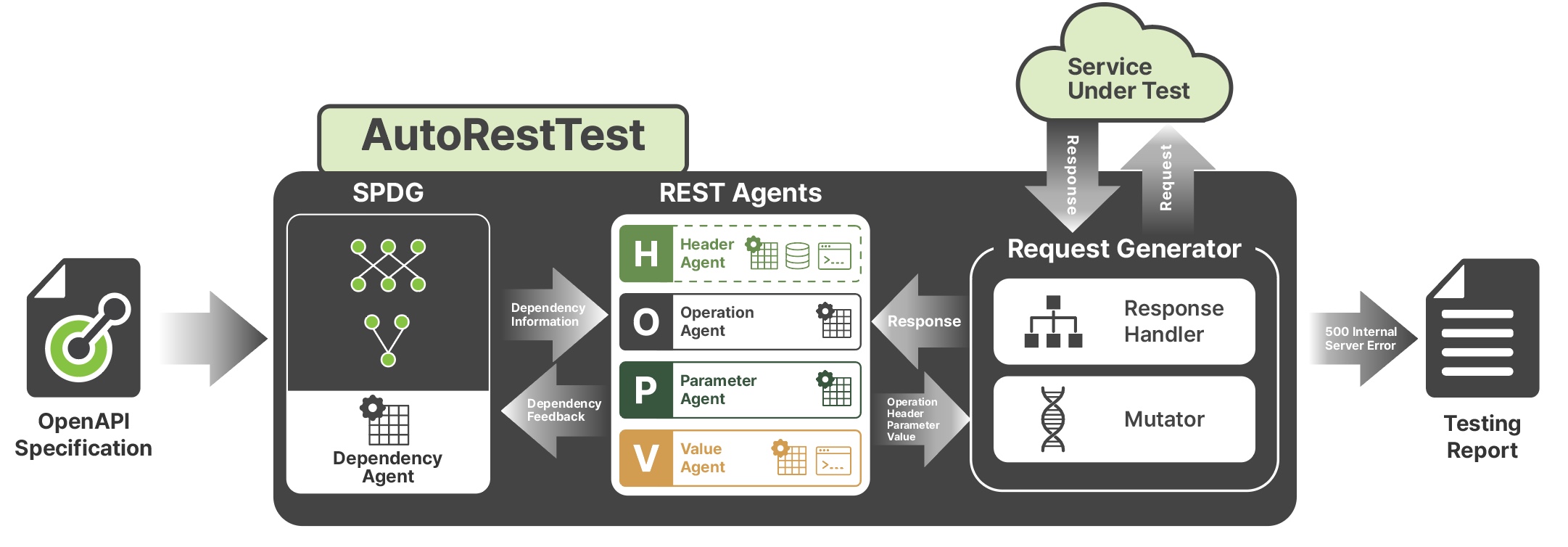}
\vspace*{-15pt}
\caption{Overview of AutoRestTest.}
\vspace*{-10pt}
\label{fig:overview}
\end{figure*}

Figure~\ref{fig:overview} illustrates the AutoRestTest architecture, which consists of three primary modules: the SPDG, the REST agents, and the request generator. The process begins by parsing the OAS to extract endpoints as well as their request/response schemas. Using this information, the dependency agent constructs the SPDG, representing API operations as nodes and semantic similarities between operation inputs and outputs as edges. Once the graph is constructed, the operation, dependency, value, parameter, and header agents are initialized with zeroed policy tables using information from the SPDG. These tables are subsequently optimized using Q-learning~\cite{sutton2018reinforcement} to identify optimal input combinations for generating realistic requests for each endpoint. In this section, we provide a brief overview of each component; a detailed discussion of our approach is available in our research paper~\cite{kim2024multi}.

\subsection{Semantic Property Dependency Graph}

The SPDG is constructed by parsing the input OAS and assigning each API operation as a vertex. AutoRestTest then iterates through each pair of vertices, using a lightweight GloVe word-embedding model~\cite{pennington2014glove} to measure semantic similarity between the parameters, request body, and responses of the two operations corresponding to the vertices. An edge is added between vertices if the similarity value between any two items exceeds a predefined threshold (0.7 in our current implementation). For operations with no dependencies above this threshold, the top three highest similarity matches are added to the graph.

The dependency agent manages the SPDG during request generation. To do so, it communicates with the value agent to use stored parameters, request bodies, and responses from successful requests in future queries, validating the dependencies identified in the SPDG.

\subsection{REST Agents}

The REST agents consist of four specialized components, each addressing a distinct aspect of the request generation process. First, the Operation Agent selects the operation for the request. Next, the Parameter Agent determines which parameters should be included. Then, the Value Agent identifies the data source and assigns values to the selected parameters, drawing from the LLM, the dependency agent, or the default settings. Lastly, the Header Agent leverages account-related operations from the specification to supply basic token authentication headers.\footnote{The header agent is a recent addition not described in our research paper.}
\subsection{Q-Learning}

Both the SPDG and REST agent modules use Q-learning to determine the optimal actions for the operation, parameter, value, dependency, and header agents. This process involves Q-table initialization, action selection, and reward delegation.

\subsubsection{Q-Table Initialization}
A Q-table is a data structure that maps action options to expected cumulative rewards. For example, the parameter agent's Q-table lists all combinations of an operation's parameters and request body properties as potential actions, with each value initialized to zero until updated through learning.

\subsubsection{Action Selection}

Agents select actions using an epsilon-greedy strategy~\cite{sutton2018reinforcement} that balances exploration and exploitation. The exploration probability (\( f(\epsilon) = 1 - \epsilon \)) with epsilon-decay allows sufficient exploration in the early stages.

\subsubsection{Reward Delegation}

After an action, AutoRestTest's Response Handler updates the Q-table using the Bellman equation~\cite{sutton2018reinforcement}. The update process involves retrieving the current Q-value, calculating the new Q-value, and updating the Q-table. Rewards are assigned as follows:
\begin{enumerate}[label=\arabic*., left=5pt]
	\item The operation agent is rewarded for finding client (4xx) and server (5xx) errors.
	\item The other agents (value, parameter, dependency, and header) are rewarded for successes (2xx).
\end{enumerate}

\subsection{Request Generator}

The Request Generator constructs and dispatches requests using data from the previous modules. It operates in a defined sequence of steps consisting of communication, modification, and response handling. During communication, it interacts with the REST agents to determine the selected operation and the assigned parameter and header values. The modification step employs a custom mutator to randomly modify requests, potentially exposing additional server errors. This mutator probabilistically selects from options such as parameter type alterations, name mutations, media type changes, random dependency selections, and token changes to generate new values of random length. Finally, the response handling step processes responses from completed requests to determine the reward for the exploring agent, thereby refining the AutoRestTest model.

\section{Tool Usage}

To use AutoRestTest, the user must first launch the Service Under Test (SUT), exposing a URL for interaction. Next, they should configure AutoRestTest according to the SUT's requirements. Users can then use AutoRestTest's CLI to run the tool, with output data being made accessible after completion.

\subsection{Configuration}

AutoRestTest offers various configuration options to tailor each module to the specific requirements of the SUT. A centralized \textbf{configurations.py} file in the root directory of the tool streamlines the process of adjusting these settings.

\subsubsection{Specification Selection}

To select the input specification corresponding to the SUT, users must note the document location relative to the root directory in the configuration file. AutoRestTest's custom parser accepts only OAS 3.0 formatted inputs. However, users can easily transfer their outdated Swagger 2.0 files using the public Swagger Converter. It is essential to ensure that the URL supplied in the specification matches the exposed URL of the SUT.

\subsubsection{Large Language Model Engine and Parameters}

To accommodate varying budgets and larger services with deeply nested objects that require extensive context windows, AutoRestTest allows users to select the Large Language Model (LLM) engine and configure parameters for its value agent from OpenAI's fleet. Specifically, AutoRestTest supports all recent models with built-in price tracking for GPT-4o, GPT-4o mini, o1, and o1-mini (for cost transparency).

The LLM temperature parameter, adjustable in AutoRestTest, influences output diversity and determinism. The default setting of \( 0.7 \) balances accuracy and creativity, with higher values (\( >1\)) producing more diverse outputs and lower values (\( \approx 0 \)) and yielding more deterministic results.

\subsubsection{Caching}

To avoid redundant use of the LLM, AutoRestTest incorporates optional local caching through Python's \textbf{shelve} object persistence library. While caching is enabled, the SPDG and LLM-generated Q-tables are stored in a database file. Subsequent executions attempt to use these cached values, significantly reducing testing costs. However, users should disable caching when making changes to the graph or Q-table to allow regeneration of the database files.

\subsubsection{Q-Learning Parameters}

AutoRestTest employs the value decomposition approach to Q-learning~\cite{sunehag2017value}, which integrates a \textit{learning rate} and a \textit{discount factor} to ensure Q-table convergence. By default, AutoRestTest uses values \( 0.1 \) and \( 0.9 \) for the learning rate and discount factor, respectively~\cite{kim2023reinforcement}. Users can adjust these parameters to influence the convergence speed and agent behavior. Basically, a higher learning rate would result in more drastic updates of the Q-table values, potentially increasing convergence speed but lowering accuracy; a lower discount rate would diminish the importance of later requests, heavily emphasizing the initial queries.

\subsubsection{Request Generator Modifications}

Two configurable parameters govern the request generation process: the time duration and the mutation rate. Given the complexity of the agent design and the large Q-tables, AutoRestTest benefits from an extended execution period to ensure adequate request diversity and convergence. In addition, increasing the mutation rate broadens the range of generated requests, while potentially slowing Q-value convergence.

\subsection{Command Line Interface}

Users can operate and interact with AutoRestTest through its CLI once the program has been configured. To begin, users should either install the requirements listed in the \textbf{requirements.txt} file or enable the Conda environment within the \textbf{auto-rest-test.yaml} file. 

The CLI regularly updates users on AutoRestTest's progress. Major milestones include the creation of the SPDG, the instantiation of the Q-learning policy tables, and the commencement of the request generation process. Given the complexity of each step, the CLI provides intermediary messages between these milestones. If unexpected errors occur, such as issues with caching, AutoRestTest notifies the user through the CLI before handling the issue and continuing.

During the request generation and Q-learning phases, the CLI outputs information about operation coverage and time elapsed. This includes the number of unique server errors identified and successful operations processed, the distribution of status codes, and the percentage of time elapsed. This continuous output allows users to quickly assess AutoRestTest's efficiency and patterns in error identification over time.

\lstset{style=codingstyle}
\begin{figure}
\begin{lstlisting}[
    language=Python,
    caption={AutoRestTest Report Output for the Market Tool},
    label={lst:autoresttest_output},
    basicstyle=\scriptsize
]
{
  "Title": "Report for 'Api Documentation' (market2)",
  "Duration": "300 seconds",
  "Total Requests Sent": 3991,
  "Status Code Distribution": {
    "500": 149,
    "401": 3509,
    "200": 99,
    "406": 215,
    "404": 19
  },
  "Number of Total Operations": 13,
  "Number of Successfully Processed Operations": 3,
  "Percentage of Successfully Processed Operations": "23.08%",
  "Number of Unique Server Errors": 104,
  "Operations with Server Errors": {
    "addItemUsingPUT": 31,
    "createCustomerUsingPOST": 45,
    "updateContactsUsingPUT": 21,
    "getProductUsingGET": 21,
    "getOrderUsingGET": 12,
    "setDeliveryUsingPUT": 6,
    "payByCardUsingPOST": 13
  }
}
\end{lstlisting}
\vspace{-4pt}
\end{figure}

\vspace{-4pt}
\subsection{Report Generation}

Upon completion, AutoRestTest compiles comprehensive data from exercising the SUT into a sequence of files for user access and evaluation. These files are available in the \textbf{data/} folder of the specified root directory:
\begin{itemize}[left=5pt]
	\item \textbf{report.json} summarizes AutoRestTest's findings, including status code distributions, total successful operations, and unique server errors. Listing~\ref{lst:autoresttest_output} shows an example report.
	\item \textbf{server\_errors.json} stores details of every request that resulted in server errors (5xx) for reproducibility.
	\item \textbf{operation\_status\_codes.json} contains the distribution of status codes for each operation.
	\item \textbf{successful\_parameters.json} lists parameter assignments that returned successful (2xx) responses by operation.
	\item \textbf{successful\_bodies.json} provides request body properties that returned successful (2xx) responses by operation.
	\item \textbf{successful\_primitives.json} contains request bodies with no associated properties that returned successful (2xx) responses by operation.
	\item \textbf{q\_tables.json} presents the converged Q-table values for each agent across all operations.
\end{itemize}

By evaluating these files, users can identify both strengths and weaknesses in a given API. The parameter agent's Q-table indicates successful parameter combinations and inter-parameter dependencies. The dependency agent exposes relationships between parameters across operations. The operation status code distribution visualizes which operations are comprehensive and easy to process. Finally, analyzing server errors can help users improve the reliability of their service.

\section{Preliminary Results}

We evaluated the performance of AutoRestTest alongside the four state-of-the-art REST API testing tools used in the ARAT-RL study: RESTler~\cite{atlidakis2019restler} (v9.2.4), EvoMaster~\cite{arcuri2019restful} (v3.0.0), ARAT-RL~\cite{kim2023reinforcement} (v0.1), and MoRest~\cite{morest} (obtained directly from the authors). All tools were tested against four real-world RESTful services included in a recent study~\cite{kim2023enhancing}, namely FDIC, OMDb, OhSome, and Spotify. To quantify effectiveness, we measured the number of successfully processed operations (2xx status codes) within a one-hour testing window, a preferred metric for comparing REST API testing tools~\cite{golmohammadi2023survey}. As shown in Table~\ref{tab:processed_operations}, our approach covered 26 unique operations across these services, outperforming ARAT-RL (12), EvoMaster (11), MoRest (11), and RESTler (10).

Notably, AutoRestTest was the only tool able to generate 2xx responses for the OhSome service, one of the most challenging RESTful services considered; while other tools were only able to trigger 4xx status codes, AutoRestTest successfully processed 12 operations in the service. Our tool was also the strongest performer on the Spotify service. Moreover, in an evaluation of internal server errors, AutoRestTest was the only tool to detect a 5xx status code on the Spotify service. We have reported the error and are awaiting a response from the developers. These results provide initial, yet clear evidence that AutoRestTest is effective in testing real-world RESTful APIs, including for complex services.

\begin{table}[t]
\centering
\caption{Number of operations exercised.}
\vspace*{-8pt}
\resizebox{\columnwidth}{!}{%
\begin{tabular}{lccccc}
\toprule
& \textbf{AutoRestTest} & ARAT-RL & EvoMaster & MoRest & RESTler \\
\midrule
FDIC & 6 & 6 & 6 & 6 & 6 \\
OMDb & 1 & 1 & 1 & 1 & 1 \\
OhSome & \textbf{12} & 0 & 0 & 0 & 0 \\
Spotify & \textbf{7} & 5 & 4 & 4 & 3 \\
\midrule
Total & \textbf{26} & 12 & 11 & 11 & 10 \\
\bottomrule
\end{tabular}
}
\label{tab:processed_operations}
\end{table}

\section{Conclusion}

We introduced AutoRestTest, a new tool that combines a SPDG, LLMs, and MARL to effectively test REST APIs. We described our approach, demonstrating how AutoRestTest addresses key challenges in API testing through advanced dependency modeling and intelligent request generation. Additionally, we conducted a preliminary study that shows the effectiveness of our tool in practice. We provided a practical demonstration of the tool’s usage and made the artifact available for further evaluation and replication~\cite{artifact}.

\section*{Acknowledgments}
\begin{small}
  This work was partially supported by 
  NSF, under grant CCF-0725202 and
  DOE, under contract DE-FOA-0002460,
  and gifts from Facebook, Google, IBM Research, and Microsoft Research.
\end{small}

\balance
\bibliographystyle{IEEEtran}
\bibliography{paper}

\end{document}